\documentclass[useAMS,usenatbib]{mn2e}
\usepackage{amssymb}
\usepackage{epsfig}

\newcommand{\iue}{{\it IUE}}
\newcommand{\hst}{{\it HST}}
\newcommand{\kms}{km s$^{-1}$}
\newcommand{\teff}{T$_{\rm eff}$}

\title[AW Per]
{The Angular Separation of the Components of the Cepheid AW Per}

\author[D. Massa and N. R. Evans]{D. Massa$^{1}$\thanks{E-mail:
massa@derckmassa.net; evans@head-cfa.harvard.edu}
\thanks{Based on observations with the NASA/ESA 
Hubble Space Telescope, obtained at the Space Telescope Science Institute, 
which is operated by the Association of Universities for Research in 
Astronomy, Inc. under NASA contract No. NAS5-26555.} and N. R.
Evans$^{2}$
\footnotemark[1]\\
$^{1}$SGT, Inc. NASA's GSFC, Code 665, Greenbelt, MD 20771, USA\\
$^{2}$Center for Astrophysics, 60 Garden St., MS 4, Cambridge, MA 02138}
\begin{document}

\date{Accepted 2007 December 15. Received 2007 December 14; in original 
form 2007 August 13}

\pagerange{\pageref{firstpage}--\pageref{lastpage}} \pubyear{2007}

\maketitle

\label{firstpage}

\begin{abstract}
The 6.4 day classical Cepheid AW Per is a spectroscopic binary with a 
period of 40 years.  Analyzing the centroids of {\it HST}/STIS spectra 
obtained in November 2001, we have determined the angular separation of the 
binary system.  Although we currently have spatially resolved data for a 
single epoch in the orbit, the success of our approach opens the 
possibility of determining the inclination, $\sin i$, for the system if the 
measurements are repeated at additional epochs.  Since the system is 
potentially a double lined spectroscopic binary, the combination of 
spectroscopic orbits for both components and the visual orbit would give 
the distance to the system and the masses of its components, thereby 
providing a direct measurement of a Cepheid mass.
\end{abstract}


\section{INTRODUCTION}

Cepheids are important stars in many respects, most notably for their 
roles as fundamental rungs on the cosmic distance ladder 
and the challenges their structure pose to stellar interiors modelling. 
The use of Cepheids as primary extragalactic distance indicators makes a
quantitative understanding of their properties extremely valuable.  
While the Magellanic Clouds are perhaps the best laboratory to study 
cepheids, the dependence of the period--luminosity relation on metalicity 
is still not fully understood (Romaniello et al.\ 2005).  Consequently, 
accurate distances (absolute magnitudes) to Galactic cepheids are needed 
to fully understand and quantify this dependence and to apply cepheid 
scale to more metal rich spiral galaxy stars which are more commonly used 
in extragalactic distance determinations.  

Cepheids also present important tests for interiors calculations since, as 
evolved stars, their structure is dictated by their evolutionary history.  
In addition, the models must predict the puslational properties of 
cepheids, making the modelling especially challenging.  This complexity is 
codified in the term ``the Cepheid mass problem''.  Forty years ago, when 
the first hydrodynamic pulsation calculations were made, it was realized 
that the mass could be derived by either matching the Herzsprung 
progression of secondary maxima or by a parameterization of the 
pulsation constant.  These masses were as much as a factor of two smaller 
than evolutionary calculations.  A reconciliation was recently achieved 
from re-evaluation of the interior opacities (see Simon, 1990, for a 
summary).  We see, therefore, that in addition to absolute magnitudes, 
obtaining accurate Cepheid masses is also important.

If we can determine the angular separations of binary systems containing 
a Cepheid, which are double lined spectroscopic binaries (SB2s), then the 
{\em distances and masses} of the Cepheids can be derived from basic 
physics.  Because of the central roles of Cepheids in fundamental 
astrophysics, it is important to have such direct measurements.  While 
several Cepheid distances have been measured directly by the 
{\it Hipparcos}\/ satellite, the quality of these measurements was only 
sufficient for statistical considerations (e.g., Groenewegen \& Oudmaijer 
2000).  More recently, a large campaign using the Fine Guidance Sensor on 
\hst\/ has begun to yield accurate distances to single Cepheids (Benedict 
et al.\ 2002).  However, to date the mass of only one cepheid, Polaris, 
has been directly determined from fundamental observations (Evans, et al.\ 
2007).

\begin{figure}
\centerline{\hbox{
\epsfxsize=3.5truein
\epsffile{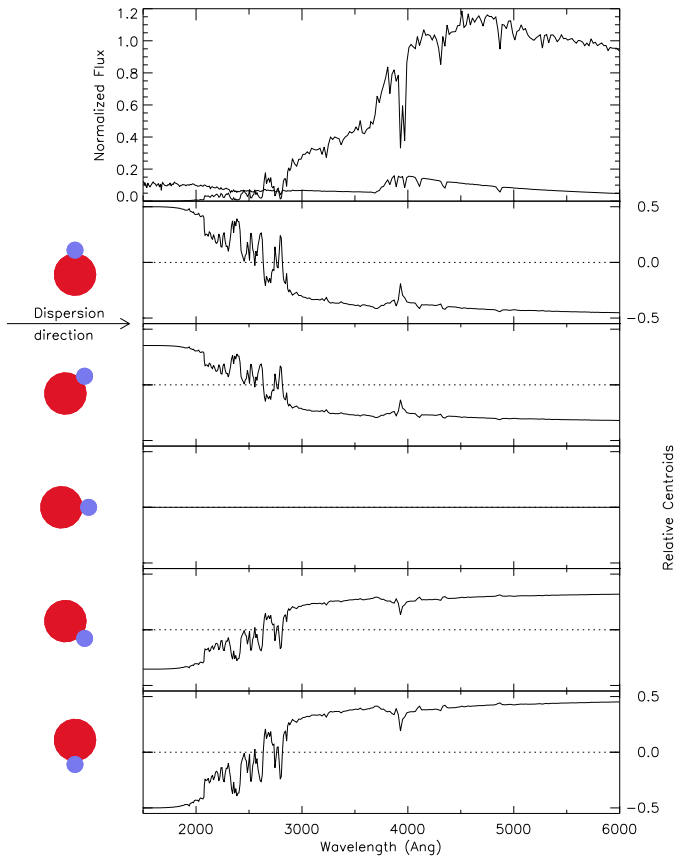}
}}
\caption{Kurucz models for a typical Cepheid (large/red) + hot star 
(small/blue) binary.  The top panel shows how the secondary is roughly 10 
times fainter in the optical, making the system extremely difficult to 
resolve from the ground.  On the other hand, the secondary dominates the 
flux from the system in the UV.  The remaining 5 panels demonstrate how the 
wavelength dependence of the spectrum centroid changes with orientation of 
the axis of the binary relative to the dispersion for 5 different 
orientations, shown to the left of each panel.  Notice that in the spectral 
region accessible from the ground, the centroid shifts by less than 10\% of 
the full separation. The ``cross-over'' point is not reached until $\lambda 
\sim $ 2500\AA.  A color version of the figure is available in the 
electronic version of the paper.}
\label{cartoon}
\end{figure}

Although SB2s containing a Cepheid and an A or B star are common (see, 
Evans 1995), these stars are difficult to resolve in the optical.  This 
is because of the inevitable, enormous magnitude differences of the 
components in the optical, which result from massive stars evolving 
toward cooler temperatures at nearly constant luminosity.  The top panel 
Figure~\ref{cartoon} shows a typical example of a Cepheid~+~B~star binary, 
and the contrast between the primary and secondary throughout the optical 
and IR is obvious.  

%%%%%%%%%%%%%%%%%%%%%%%%%%%%%%% Figure 1 %%%%%%%%%%%%%%%%%%%%%%%%%%%%%%%%%
%%%%%%%%%%%%%%%%%%%%%%%%%%%%%%%%%%%%%%%%%%%%%%%%%%%%%%%%%%%%%%%%%%%%%%%%%

Thus, while the measurement of a Cepheid mass by directly imaging a double 
lined spectroscopic binary with a Cepheid primary and an A or B star 
secondary has been a long-sought goal, ground-based studies have not, as 
yet, been able to accomplish this (even though they have been able to 
resolve the stellar disks of some Cepheids, e.g., Kervella et al.\ 2004, and 
references therein).  As a result, indirect methods have been developed to 
determine the masses of Cepheids.  The most popular of these uses a 
combination of UV and optical spectroscopy to obtain radial velocity curves 
for both components.  Then the UV spectral energy distribution (SED) of the 
hot secondary is used to obtain its temperature.  Finally, the mass -- 
temperature relation for main sequence A or B stars is used to {\it 
infer}\/ the mass of the secondary and, thus, (since the system is an SB2) 
the mass of the primary.  This approach has been applied to several systems 
(SU~Cyg, S~Mus and V350~Sgr), using \iue\/ or \hst\/ spectra to determine 
the radial velocity curves and SEDs of the secondaries (Evans, et al., 
1998).  The masses obtained by this approach agree, on average, with the 
mass-luminosity predictions from evolutionary calculations with moderate 
convective overshoot (e.g. Schaller, et al.\, 1992).  However, this 
approach requires an exact understanding of the evolutionary phase of 
the hot secondary and relies on its spectroscopic parallax to determine the 
distances to the systems.  Clearly, a direct measurement of the masses of 
both components is more desirable.  

In this paper, we describe how we used the Space Telescope Imaging 
Spectrograph (STIS) on {\it HST}\/ to resolve a the potential SB2 Cepheid 
binary AW~Per using an approach we call ``cross-dispersion imaging''.  
AW~Per is a 6.4~day Cepheid which is in a roughly 40~year orbit with its 
hot secondary (Evans et al.\ 2000).  Evans (1989) studied the system and 
determined that the secondary is a main sequence B7-8 star and that the 
color excess of the system is $E(B-V) = 0.52$.  The \teff\/ of the 
secondary is expected to be $\sim 12000$K (Evans 1994).  

The remainder of the paper is organized as follows: \S\ref{approach} 
provides an overview the approach used to ``resolve'' the binary, 
\S\ref{observations} describes the observations, \S\ref{data} gives the 
data analysis, \S\ref{analysis} details the analysis of the observations, 
\S\ref{results} presents the results, \S\ref{discussion} discusses the 
results and their implications, and \S\ref{summary} summarizes the 
findings.   

\section{THE APPROACH{\it (Cross-Dispersion Imaging)}}\label{approach}

\subsection{Basic Principles} 

Massa \& Endal (1987) describe how combining imaging and spectroscopy can 
dramatically increase the effective ``resolving power'' of an instrument.  
Specifically, they showed how the wavelength dependence of the centroid of 
a spectrum can determine the angular separation of an unresolved binary 
whose components have distinctly different spectra.  The basic concept of 
this approach is quite simple.  It is based on an idea put forth by Beckers 
(1982) and has been independently discovered by a number of others (see, 
e.g., Porter et al.\ 2004, and references therein).  

Like all cross-dispersion imaging techniques, some sort of a model is 
required to interpret the observations.  This model might be extremely 
simple, as in the case of a binary where one assumes that the system is 
composed of {\em exactly} two stars, and that one contributes {\em all} 
of the flux at one wavelength and the other contributes all of the flux 
at another wavelength.  This crude model would be sufficient to 
``resolve'' the binary from the properties of its spectrum.

Consider the image of a highly unresolved binary system.  To first order, 
the image of the combined light from the system is indistinguishable from 
a point source.  However, the position of an image at any given wavelength 
will be displaced toward the location of the binary component which 
contributes most of the light that wavelength.  In principle, one could 
obtain images at several different wavelengths and determine how the 
centers of the images shift from one exposure to the next.  Analysis of 
this set of data (along with a model for the flux ratios in each band) 
would then determine the separation of the two components (Becker 1982).  
The drawback of this direct approach is that all of the exposures would 
have to be obtained using different optical elements, making alignment at 
the sub-pixel level effectively impossible.  Instead, Massa \& Endal 
(1987) show that tracking the centroid of the spectrum of the binary has 
the same effect.  Furthermore, because all of the position measurements 
(the centroid of the spectrum at each wavelength) are obtained at one 
time, this method is more efficient and the measurements are differential 
in nature, freeing them from several sources of systematic error.  

To make these notions quantitative, let $x$ and $y$ be the angular 
coordinates on the detector which are parallel and perpendicular to the 
wavelength dispersion.  Therefore, the wavelengths, $\lambda$, are given 
by $\lambda = \lambda(x)$.  Now, consider a binary whose components have 
an angular separation $\theta$ and photon fluxes per unit wavelength 
$N_p(\lambda)$ and $N_s(\lambda)$ for the primary and secondary, 
respectively.  Further, let $\phi$ be the position angle of the binary on 
the sky (measured c.c.\ from north toward east of a line from the primary 
to the secondary) and let $\alpha$ be a similarly measured angle between 
north and a line in the dispersion direction pointing in the direction of 
decreasing wavelength.  Thus, $\alpha$ can be varied by changing 
the orientation of the telescope.  With these definitions, the wavelength 
dependence of the centroid of the spectrum of a single observation of a 
binary is 
\begin{eqnarray} 
  y(\lambda) & =& \frac{\Delta y}{1+N_p(\lambda)/N_s(\lambda)}+Const. 
    \; \; \; \; \; {\rm where} \label{rlam} \\
  \Delta y & = & \theta \sin(\phi - \alpha) \label{solve}
\end{eqnarray}
(see the appendix).  Thus, if $N_p(\lambda)/N_s(\lambda)$ is known, then 
measurements at two or more orientations ($\alpha$'s) enables one to 
determine $\theta$ and $\phi$, the separation and position angle of the 
binary.  Note that if the spectral energy distributions (SEDs) of the 
components are vastly different, then the position of the centroid shifts 
from one to the other, depending upon which star dominates the flux at each 
wavelength.  On the other hand, if the binary components have identical 
SEDs, then no spatial information can be gained from the centroid 
positions.  

Figure \ref{cartoon} is a cartoon depicting how the centroid of the 
spectrum of a binary star, whose components have very different effective 
temperatures, is influenced by the relative energy distributions of the two 
components and the orientation of the binary relative to the dispersion 
direction of a spectrograph.  In this case, the centroid shifts from the 
cool component at long wavelengths to the hot component at short 
wavelengths.  We define the {\em cross-over wavelength} as that wavelength 
were each binary component contributes equally to the flux.  For Cepheid 
binaries, this wavelength is typically in the near UV ($\sim 3000$\AA\ for 
the case shown).  In order to infer spatial information from the centroids, 
it is desirable to span as large a wavelength baseline as possible, to 
maximize the deflections in the centroid positions.  The best case would be 
to cover a large enough wavelength range with a single setting, so that one 
end of the spectrum is totally dominated by one star and the other end is 
dominated by the other.  If this is not practical, a wavelength band 
centered on the cross-over wavelength and covering a baseline large enough 
to experience more than a 50\% centroid deflection is adequate.  However, 
in this case, one needs an estimate of the SEDs of the two binary 
components in order to extract the angular separation.  Note that if the 
absolute flux calibration of the instrument is well-determined, then the 
flux observations can provide additional information which can be 
incorporated into the determination of the angular separation (see 
\S\ref{analysis}).  

Finally, to unambiguously determine the separation and position angle of 
the binary, two or more observations are required in order to solve 
eq.~(\ref{solve}) for $\theta$ and $\phi$ in terms of the measured 
quantities $\Delta y^{(n)}$ and $\alpha^{(n)}$, for $n \geq 2$.
  
\begin{figure}
\centerline{\hbox{\epsfxsize=3.5truein\epsffile{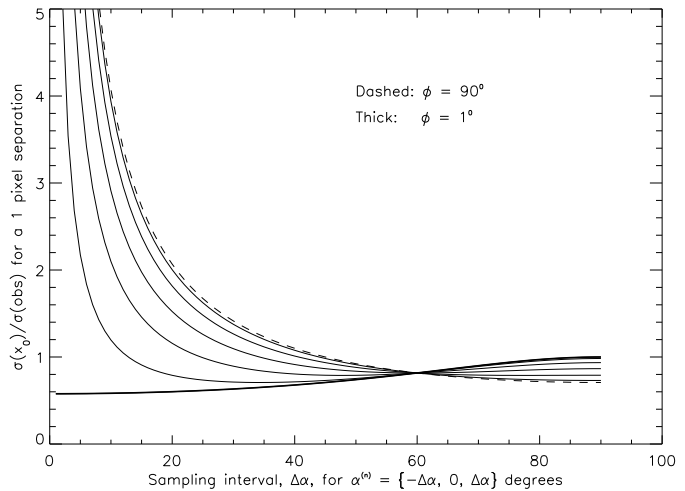} }}

\caption{Relative error in the angular separation of a binary determined 
from fitting a cosine curve to measurements obtained at three orientations, 
$\{-\Delta \alpha, 0, +\Delta \alpha\}$ versus $\Delta \alpha$ (abscissa)
over the interval $\Delta \alpha = 1 \rightarrow 90^{\circ}$.  The different 
curves are for different values of the orientation of the system on the sky, 
$\phi$, between $\phi = 1 \rightarrow 90^{\circ}$.  
}\label{sampling}
\end{figure}

The final error associated with the angular separation and the position 
angle measurements depends upon the band pass of the observation, the 
signal-to-noise of the data (discussed in the next section), the 
number of independent orientations obtained and the relation between 
the these angles and $\phi$.  We have examined the relative error for 
sampling three orientations, $\alpha^{(n)} = \{-\Delta \alpha, 0, 
+\Delta \alpha\}$, for position angles between 1 and $90^{\circ}$.  
Figure~\ref{sampling} demonstrates how the relative accuracy of the 
observations changes as a function of sampling interval, $\Delta 
\alpha$, and relative orientations, $\phi$, for this case.  For most 
orientations, any sampling with $\Delta \alpha \gtrsim 30^{\circ}$ 
provides comparable accuracy.

%%%%%%%%%%%%%%%%%%%%%%%%%%%%%%%%
The approximations developed in this section are only valid in the 
sub-Rayleigh regime.  Once the sources are resolved at any wavelength, 
the entire image must be modeled using a an accurate representation of 
the point spread function as well as the fluxes of the two objects.  
%%%%%%%%%%%%%%%%%%%%%%%%%%%%%%%%

%%%%%%%%%%%%%%%%%%%%%%%%%%%%%%% Figure 2 %%%%%%%%%%%%%%%%%%%%%%%%%%%%%%%%%
%%%%%%%%%%%%%%%%%%%%%%%%%%%%%%%%%%%%%%%%%%%%%%%%%%%%%%%%%%%%%%%%%%%%%%%%%%%

\subsection{Exposure Times and Random Errors} 

The counts needed to centroid to an accuracy $\sigma [y(\lambda )]$ 
can be estimated for an instrument whose spread function perpendicular to 
the dispersion is a Gaussian with $FWHM=\xi$.  A single count is equivalent 
to one estimate of the center of the spectrum drawn from a sample with an 
RMS dispersion $\sigma =\xi /\sqrt{8\ln 2} =0.42\xi$.  Therefore, $N$ 
samples (counts) determine the centroid to an accuracy of 

\begin{equation}
\sigma [y(\lambda )]=\frac{0.42\xi}{\sqrt{N(\lambda )}}.
\label{times}
\end{equation}

Equation (\ref{times}) gives the counts needed over a wavelength band to 
obtain the desired accuracy.  The {\it FWHM}\/ of the STIS PSF varies from 
$\sim 0.05 - 0.07$\arcsec\/ (depending on wavelength) and the minimum number 
of counts obtained in one 10 min observation over a spectral resolution 
element (2 pixels) was $4000$, and we obtained 3 of these.  Therefore, the 
{\em poorest}\/ precision we can expect based upon simple sampling 
arguments is $\sim 3 \times 10^{-4}$\arcsec, and this is for a single 
resolution element.  In all, there are 512 independent resolution elements 
which will be combined to determine a single measurement of $\Delta y$ 
through the use of eq.~(\ref{rlam}).  Therefore, random noise in the 
angular separation determinations should be $\lesssim 10^{-4}$\arcsec, and 
not a limiting factor for our observations.  However, as is typical for 
most observations, we shall see that systematic effects will dominate the 
error budget (see, \S\ref{data}).

\section{THE OBSERVATIONS}\label{observations}

As can be seen from the top panel of Figure \ref{cartoon}, a broad 
wavelength baseline is needed to optimize the extraction process.  
Furthermore, good spectral resolution is also advantageous, since spectral 
features provide additional constraints.  Consequently, we employed the 
STIS on {\it HST}\/ to obtain high spatial resolution, excellent wavelength 
coverage and good spectral resolution.  We used the STIS NUV-MAMA detector 
together with its G230L grating, since this combination provided good 
coverage ($1600 \leq \lambda \leq 3160$~\AA) of the expected cross-over 
point (see, Kim Quijano, J., et al. 2003).  

Spectra were obtained at three distinct roll angles (see, Table~\ref{log}) 
which differ by $\sim \pm 20^\circ$.  Although rolls of $\pm 60^\circ$ 
would be optimal, we were limited to smaller rolls by \hst\/ restrictions 
for objects at the declination of AW~Per.  Although not optimal, 
Figure~\ref{sampling} shows that this restricted range does not sacrifice 
very much in theoretical accuracy.  After a standard STIS target 
acquisition, which centers the binary within a 0.1\arcsec\/ aperture, we 
obtained the science observations through the 25MAMA aperture, which 
provides slitless spectra of the binary.  At each roll, we offset the star 
by $\pm 0.1$\arcsec\/ and obtained additional science exposures.  This 
procedure allows us to characterize localized distortions in the detector.  
It is also useful for determining the sensitivity of the observations to 
their position on the detector, since each spectrum is sampled differently 
by the pixel lattice.  Since the spectrum was repositioned to within 2 
pixels ($<0.05$\arcsec) after each roll, the dispersion of measurements 
obtained at the $\pm 0.1$\arcsec\/ offsets should provide a good 
characterization of the errors that result from all of the changes 
encountered in the positioning of the spectrum.  The reproducibility of 
these observations also provides a more realistic measurements of the 
centroiding errors than those based on simple signal-to-noise 
considerations.  As a result of our observing strategy, we obtained 3 
observations at each of 3 rolls, for a total of 9 spectra, with exposure 
times of roughly 10~min each.  

\begin{table*}
 \begin{minipage}{120mm}
  \caption{Observation log\label{log}}
\begin{tabular}{lcrrcrrr}
\hline
Obs ID & Off Set & Roll & Obs Date & Exp. Time & Phase\footnote{Phase, 
$V$ and $(B-V)$ are derived from sources in the literature, as discussed 
in the text.} & $V$ &$(B-V)$ \\
      & arc sec & Deg. & MJD - 52235 & Min. & $\Phi$ & Mag. & Mag. \\ \hline
o6f104010  & $+0.0$ & 175.526 & 0.34765625 & 10.0 & 0.906 & 7.40 & 1.02 \\ 
o6f104020  & $+0.1$ & 175.526 & 0.35546875 & 10.0 & 0.907 & 7.39 & 1.01 \\
o6f104020  & $-0.1$ & 175.526 & 0.36328125 & 11.4 & 0.909 & 7.38 & 1.01 \\
o6f105010  & $+0.0$ & 205.000 & 0.41406250 & 10.0 & 0.916 & 7.34 & 1.00 \\ 
o6f105020  & $+0.1$ & 205.000 & 0.42187500 & 10.0 & 0.918 & 7.33 & 0.99 \\ 
o6f105030  & $-0.1$ & 205.000 & 0.42968750 & 11.4 & 0.919 & 7.32 & 0.99 \\
o6f106010  & $+0.0$ & 146.526 & 0.48046875 & 10.0 & 0.927 & 7.27 & 0.97 \\ 
o6f106020  & $+0.1$ & 146.526 & 0.48828125 & 10.0 & 0.928 & 7.26 & 0.97 \\ 
o6f106030  & $-0.1$ & 146.526 & 0.49609375 & 11.4 & 0.929 & 7.26 & 0.97 \\ 
\hline
\end{tabular}
\end{minipage}
\end{table*}

The orientations mentioned above are measured with respect to the STIS 
coordinate system, which we define as the $x_0 - y_0$ system.  In this 
system, the dispersion direction (from red to blue) makes an angle 
(measured in the c.c.\ direction) of $-1.4^\circ$ with the $x_0$ axis.  

\section{DATA REDUCTION}\label{data} % 0.0247 \arcsec per pixel

\subsection{Centroids}

The first step in the reduction process was to extract the centroids.  This 
presents a problem, since the STIS detector does not oversample the \hst\/ 
PSF.  However, since (as will be explained shortly) only relative centroids 
will be needed, we can accept some level of bias in the extraction process, 
as long as it is consistent.  This is because the ultimate measurements 
will be differences of the centroids, which will cancel small, uniform 
biases introduced in the extraction process.  

We used three separate approaches to extract the centroids, $y(\lambda)$, 
from the {\em raw} images.  We chose to analyze the raw images (in their 
native ``highres'' $2048\times2048$ format) because initial experimentation 
showed that the geometrically corrected images did little to improve the 
relative positions of the centroids over the a range of 10 pixels or less 
(which are the scales important to us).  Thus it was felt best to avoid 
the inevitable smoothing which is introduced by the resampling involved 
in geometric corrections.  

The first approach we used was a simple cross-correlation technique 
relative to a set of 0.025\arcsec\/ FWHM gaussians.  The second one 
employed a standard cross-correlation technique using the cross dispersion 
profiles of a spectrum of a standard star (the wd GD71) which was observed 
at roughly the same position on the detector with the same grating.  We 
used sinc interpolation in the cross-correlation to compensate for the 
undersampling of the PSF by the MAMA detector.  Finally, we used a 
non-linear least squares fit to a set of gaussians whose FWHMs, central 
positions and amplitudes were allowed to vary at each pixel.  No 
systematic differences were found among all three approaches.  However, 
the results from the non-linearly extracted centroids produced the results 
with the lowest pixel-to-pixel scatter, and these were adopted for the 
following analysis.  

\begin{figure}
\centerline{\hbox{
\epsfxsize=3.5truein
\epsffile{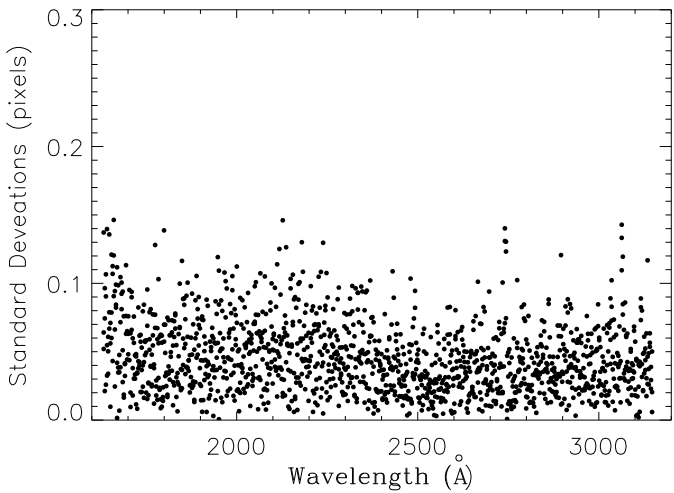}
}}
\caption{Standard deviations of the three independent spectra of AW Per 
obtained at each roll angle.  The standard deviations for each roll angle 
are over plotted.}
\label{STDS}
\end{figure}

The 3 sets of centroid measurements at each roll angle were rebinned to 
512 elements from the 2048 elements available in the raw images, and these 
were used to construct mean centroids at each roll and their standard 
deviations.  Because the centroids near the edges of the detector are 
poorly determined, of the 512 bined pixels (in the wavelength direction) 
only about 490 are well-behaved.  The standard deviations for these 490 
pixels determined for each roll angle are over plotted as a function of 
wavelength in Figure~\ref{STDS}.  The RMS means for each roll angle are 
0.027, 0.024, and 0.027 pixels or (0.67, 0.59, and 0.67 mas).  Remember, 
these are the single observation standard deviations for a single pixel, 
and there are 9 independent observations with 490 useful pixels.  Notice 
also that this scatter is significantly larger than the one expected from 
the simple signal-to-noise arguments of the previous section.  The reason 
is that the actual uncertainties are set by random differences between the 
photometric and geometric centroids of the pixels, and by localized 
geometric distortions in the detector over the range of a few pixels.  
Nevertheless, the repeatability of the centroids (to a few percent of a 
pixel) is considered quite good, and we will use this scatter to 
characterize the actual measurement errors in the centroid positions.  

%%%%%%%%%%%%%%%%%%%%%%%%%%%%%%% Figure 3 %%%%%%%%%%%%%%%%%%%%%%%%%%%%%%%%%
%%%%%%%%%%%%%%%%%%%%%%%%%%%%%%%%%%%%%%%%%%%%%%%%%%%%%%%%%%%%%%%%%%%%%%%%

Since the centroids are extracted from the raw images, they contain large 
scale geometric distortions.  Consequently, we will analyze the {\em 
relative} centroids.  To construct these, we first combine the centroids 
determined at each offset for a particular roll angle to produce a mean 
centroid, $\langle y \rangle$, at each roll.  These measurements contain 
geometric distortions and any systematic effects introduced by the centroid 
extraction technique.  However, when we analyze the differences between 
each individual mean and the grand mean of all the observations, these 
systematic affects are removed.  This is because the offsets at each roll 
are larger than the displacements from one roll to another, and the scatter 
that the former exhibit (Fig.~{\ref{STDS}) demonstrates that localized 
geometric distortions are small.  Similarly, any systematic affects that 
result from mis-matches between the actual PSF orthogonal to the dispersion 
and the gaussian used to determine the centroids will cancel out, since the 
same process is used in each case.

Finally, we must account for the fact that $y(\lambda)$ is not exactly 
perpendicular to the dispersion.  As a result, we must divide the final 
displacements that we measure by $\cos(1.4^\circ)$.

\subsection{Fluxes}

STIS fluxes were extracted from the images using the {\sc CALSTIS IDL} software 
package developed by Lindler (1998) for the STIS Instrument Definition 
Team.  In order to constrain the B star flux contribution, we also
incorporate the available \iue\/ spectra (obtained when the Cepheid 
component was near minimum light), into the analysis given in 
\S\ref{analysis}.  The \iue\/ fluxes were placed upon the \hst/STIS flux 
system using the transformations described by Massa \& Fitzpatrick (2000).  
Figure~\ref{fluxes} compares the \iue\/ and STIS spectra.  It is 
immediately clear that the \iue\/ long wavelength spectra were obtained 
when the Cepheid was near minimum light ($\Phi = 0.53$, Evans 1989), while 
the STIS observations were near maximum light (Table~\ref{log}).  
The effects of extinction are also clearly apparent, as is the fact that 
the \iue\/ fluxes are a factor of 1.146 smaller than the STIS fluxes.  
This discrepancy is a constant over the region of overlap, and its origin 
is unknown.  Consequently, we cannot be certain which set of fluxes is 
correct.  In \S\ref{results} we show that this ambiguity introduces a 
significant uncertainty into our results.  

%%%%%%%%%%%%%%%%%%%%%%%%%%%%%%% Figure 4 %%%%%%%%%%%%%%%%%%%%%%%%%%%%%%%%%
\begin{figure}
\centerline{\hbox{
\epsfxsize=3.5truein
\epsffile{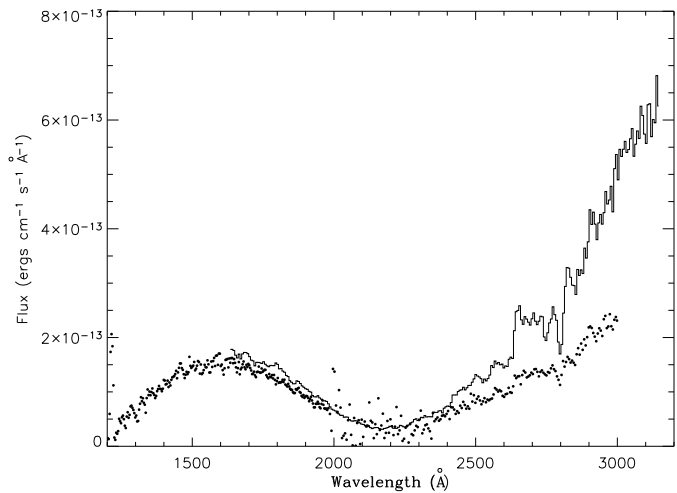}
}}
\caption{Plots of the mean STIS spectrum of AW Per (solid curve) together 
with the available \iue\/ spectra (dotted), calibrated to the \hst\/ flux 
system.}
\label{fluxes}
\end{figure}
%%%%%%%%%%%%%%%%%%%%%%%%%%%%%%%%%%%%%%%%%%%%%%%%%%%%%%%%%%%%%%%%%%%%%%%%%%

The variability of the Cepheid is clearly detectable in the STIS spectra.  
Figure~\ref{ratios} shows STIS flux ratios for the mean spectra obtained 
at the second and third roll angles divided by the first.  The time 
lapsed between the mean observations is 1.59 and 3.19 hours, respectively.  
This plot demonstrates two things.  First, the Cepheid flux changed 
significantly throughout the three \hst\/ orbits spanned by the 
observations.  Second, the flux ratios decrease with wavelength, becoming 
unity at the shortest wavelengths.  This is contrary to what is normally 
seen in single Cepheids like $\delta$~Cep (Schmidt \& Parsons 1982) where 
the flux changes typically increase with decreasing wavelength.  
Consequently, this figure shows that the flux at the shortest wavelengths 
is dominated by the B star, which does not vary.  

%%%%%%%%%%%%%%%%%%%%%%%%%%%%%%% Figure 5 %%%%%%%%%%%%%%%%%%%%%%%%%%%%%%%%%
\begin{figure}
\centerline{\hbox{
\epsfxsize=3.5truein
\epsffile{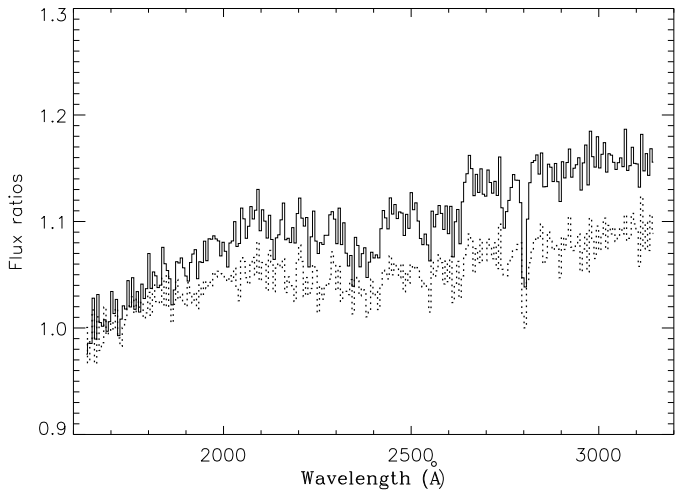}
}}
\caption{Plots of the ratios of mean STIS spectra of AW Per obtained at the 
second and third roll angles divided by the mean flux obtained at the first 
roll angle.  These plots demonstrate how the Cepheid component brightened 
over the 3.5 hour observing sequence.  Notice that the flux at the shortest 
wavelengths does not change, since it is dominated by the B star secondary.}
\label{ratios}
\end{figure}
%%%%%%%%%%%%%%%%%%%%%%%%%%%%%%%%%%%%%%%%%%%%%%%%%%%%%%%%%%%%%%%%%%%%%

The following analysis also requires the color and magnitude of the system 
the time of the observations.  We combined the data from Szabados (1980), 
Moffett \& Barnes (1984), Szabados (1991), and Kiss (1998), using the 
period and HJD for zero phase from Kiss (1998).  The combined data were 
fit with a high order polynomial, and this was used to determine the $V$ 
and $(B-V)$ photometry at the times of the STIS observations.  The 
resulting phases and photometry are listed in Table~\ref{log}.  

\section{ANALYSIS}\label{analysis}

\subsection{Overview}

Because our spectra cover a limited band-pass, we require an estimate 
for the flux ratio of the binary components in order to extract the 
wavelength dependence of the centroids.  This flux ratio is constrained, 
since it must also satisfy the observed flux of the system, which is the 
reddened, combined flux of the two binary components.  Ideally, one would 
fit the observed flux and centroid positions with a combination of single 
star spectra obtained with the same instrument and which experience the 
same reddening.  However, because there is no such library of single star 
spectra available, we used an approach which employs a model for the B 
star SED star and for the UV extinction to construct the combined flux and 
the centroids.  We  then used a non-linear least squares fitting 
procedure\footnotemark \footnotetext{We use the Markwardt non-linear {\sc 
IDL} fitting procedure, available at 
http://astrog.physics.wisc.edu/~craigm/idl/idl.html.} 
to fit the centroids and fluxes {\em simultaneously}.  This method is 
described in detail in \S~\ref{models}.

\subsection{Model Components\label{components}}

We now describe the components of the model used to fit the observations.  
In a few instances, refinements might increase the accuracy, but in the 
interest of expediency, certain effects were ignored for the first attempt.
First, we use Kurucz (1991) Atlas~9 models with updated 
metallicities\footnotemark \footnotetext{We used the the Kurucz ``preferred 
models'' available at http://kurucz.harvard.edu/.} for the B star.  We use 
only models with a micro-turbulent velocity of 2.0~\kms.  The synthetic 
photometry for the models was calibrated as in Fitzpatrick \& Massa (2005).  
We set $\log g = 4.0$ for the B star atmosphere.  The sensitivity of our 
results to this assumption is tested once a fit is achieved.  The model 
atmosphere fluxes were prepared in the manner described by Fitzpatrick \& 
Massa (2005), which is best suited to the \iue\/ fluxes.  The dust model is 
quite general.  We use the Fitzpatrick (1999) formulation of the 
Fitzpatrick \& Massa (1990) model since we need a representation of the 
near-UV extinction, and the original Fitzpatrick \& Massa (1990) 
formulation does not provide one.  Although the Fitzpatrick (1999) curve 
for the near UV is largely untested, it is reasonable and the best 
currently available.  To provide additional flexibility to the Fitzpatrick 
model, we allow the bump strength ($c_3$), the width of the 2175~\AA\ 
($\gamma$) and far-UV curvature term ($c_4$) to vary independently.  In 
this way, we can accommodate any observed extinction curve.  As a result, 
the $R_V$ parameter (the ratio of visual extinction to color excess) only 
affects the general slope of the UV extinction and the shape of the 
near-UV curve, and the wavelength dependence of the total extinction to an 
object can be expressed as, 
\begin{equation}
A_\lambda \equiv A[R_V, E(B-V), \gamma, c_3, c_4; \lambda] \;.
\end{equation}

\subsection{Details of the Fitting Procedures}\label{models}

We {\it simultaneously} fit the STIS centroids at all three roll angles and 
the \iue\/ flux from the B star.  We constrain the reddened model for the 
B star by assuming that all of the flux from the system for $\lambda \leq 
1650$~\AA\ is due to the B star.  The difference between the observed flux 
and the reddened B star model provides the Cepheid SED which is used in 
fitting the centroids.  The free parameters of the fit are: The three 
$\Delta y^{(n)}$ (displacements perpendicular to the dispersion at each 
roll angle), T$_{\rm eff}^s$ (the effective temperature of the B star 
secondary), ${\rm [m/H]}_s$ (the abundance parameter for the B star), 
$E(B-V)$ (the color excess of the system, consistent with the fluxes), 
$R_V$ (which determines the slope of the UV extinction curve), $\gamma$ 
(the width of the 2175~\AA\ bump), $c_3$ (the bump strength), and $c_4$ 
(the strength of the far UV curvature) -- 10 parameters in all.  The $V$ 
magnitude of the B star, $V_s$, is fixed by the observed flux attributed to 
the B star at $\lambda = 1650$~\AA\ and the extinction at that wavelength 
relative to $V$.  In addition to the separations, the results also yield an 
empirical, unreddened UV SED and photometry for the Cepheid.  These can 
then be and compared to models or to actual Cepheids.  Since the derived 
Cepheid flux is identical to the observed flux minus the B star flux for 
wavelengths longward of 1650~\AA, the flux in this region is fit exactly.  
The equation used to fit the centroids is: 
\begin{equation}
y(\lambda)^{(n)}  =  \Delta y^{(n)}  
                     \left[1 +\frac{N(\lambda)_{obs}^{(n)}
		   -\theta_s^2 N({\rm T}_s, \log g_s, v_t, {\rm [m/H]}; 
		   \lambda)} {\theta_s^2 N({\rm T}_s, \log g_s, v_t, 
		   {\rm [m/H]}; \lambda)}\right]^{-1} 
\end{equation}
and the unreddened flux of the Cepheid is given by
\begin{eqnarray}
N(\lambda)_p^{(n)} = & [ N(\lambda)_{obs}^{(n)} - \theta_s^2 N({\rm T}_s, 
                     \log g_s, v_t, {\rm [m/H]}; \lambda)] \nonumber \\ 
                   & \times 10^{A[R_V, E(B-V), \gamma, c_3, c_4; \lambda]} 
\end{eqnarray}
where $\theta_s$ is the angular diameter of the B star (fixed by the flux 
at 1650\AA) and $n = 1, 2, 3$ represents the observations obtained at each 
roll angle, which are means of the data for the three off-set positions.  
We cannot use a single mean for the fluxes, since significant changes in 
$V$, $(B-V)$ and the UV SED occur over the course of the observations 
(see, Table~\ref{log}, Fig.~\ref{ratios}) and must be taken into account.  
However, the data were averaged at each roll, since the time between 
off-sets was much smaller than the time between rolls.  

A major advantage of our approach is that it only relies on a Kurucz 
Atlas~9 model for the B star, and recent work by Fitzpatrick \& Massa 
(1999, 2005) has demonstrated that these provide excellent representations 
of low resolution B star SEDs.  Further, it avoids using the Atlas~9 models 
for the Cepheid component, which is desirable since the accuracy of 
Cepheid model atmospheres has not been fully tested, especially in the UV.  
This issue is addressed further in \S\ref{results}.  The disadvantage of 
our approach is that we must have extremely well calibrated fluxes, and 
we have already seen an inconsistency between the poorly exposed \iue\/ 
fluxes and the STIS data.  

\subsection{Determining the Separations}

The final step in the analysis is to fit the angular separations derived 
at each roll angle to a sine curve whose phase and amplitude are related 
to the position angle and separation of the binary (eq.~\ref{solve}).  
The amplitude of the curve is the full separation of the system and the 
phase is the position angle of the system on the sky.  The abscissa of the 
plot is the position angle in the $x-y$ system, which is equal to the 
values listed in Table~\ref{log} minus $1.41^{\circ}$ (which accounts for 
the rotation to align the spectra with the $y$ axis).  Figure~\ref{angles} 
shows the definitions of the different angles used in the analysis, and 
their relations to one another.

%%%%%%%%%%%%%%%%%%%%%%%%%%%%%%% Figure 6 %%%%%%%%%%%%%%%%%%%%%%%%%%%%%%%%%
\begin{figure}
\centerline{\hbox{
\epsfxsize=3.5truein
\epsffile{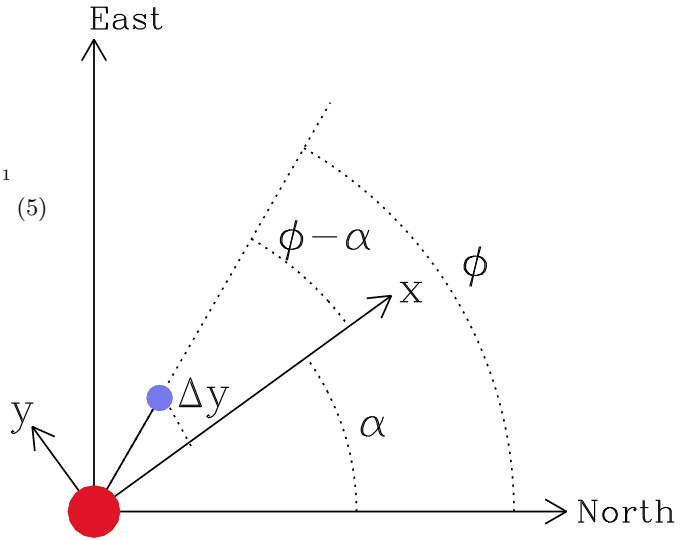}             
}}
\caption{Diagram showing the definitions of the different angles and 
coordinate systems used in the analysis, and their relations to one 
another.  The position angle on the sky of the binary angle, $\phi$, is 
defined as the angle measured the c.c.\ from north to east, with the 
primary at the origin.  The $x-y$ system is the standard STIS coordinate 
system, with $x$ parallel to the dispersion (increasing in the direction of 
increasing wavelength) and $y$ perpendicular to it.  The angle $\alpha$ 
(also measured the c.c.\ from north to east) is defined as the angle 
between North and $x$ for a given telescope orientation.  Thus, $\phi 
-\alpha$ is the angle between the dispersion and a line connecting 
the binary components and $\Delta y  = \theta \sin(\phi - \alpha)$ is the 
displacement of the two spectra of the binary perpendicular to the 
dispersion.  If $\phi - \alpha = 0$ or  $\pm 180^\circ$, then $\Delta y = 
0$.} 
\label{angles}
\end{figure}
%%%%%%%%%%%%%%%%%%%%%%%%%%%%%%%%%%%%%%%%%%%%%%%%%%%%%%%%%%%%%%%%%%%%%%%%%

\subsection{Weights}

The non-linear least squares involves fitting an array which consists 3 
sets of centroids and the \iue\/ fluxes all at once.  To perform the fit, 
we must provide errors for the different components of this array.  The 
measurement errors affecting the centroids were obtained from the standard 
deviations of the three independent sets of measurements obtained at each 
offset position.  For the \iue\/ data, we used the error vector which 
accompanies the MXLO fluxes (see, Nicholes \& Linsky 1996).  

\section{RESULTS}\label{results}

In fitting the data, we assumed a microturbulent velocity of 2.0~\kms, 
which is typical for main sequence B stars (e.g., Fitzpatrick \& Massa 
2005).  Because the B star is overwhelmed by the Cepheid in the optical 
and near-UV, we do not have access to the classical $\log g$ diagnostics 
for B stars, namely, the Balmer jump and Balmer lines.  Consequently, we 
fixed the surface gravity at 4.0, again typical for main sequence B stars.  
We allowed the abundance parameter, ${\rm [m/H]}_s$, and the effective 
temperature of the B star to be optimized by the least squares routine, 
along with the $\Delta y$'s and the extinction parameters. In addition, we 
assumed that the \iue\/ fluxes were correct (so the STIS fluxes were 
divided by 1.146 to make them agree with the \iue\/ data).  In applying 
our model, we also assume that all of the STIS flux in a 30~\AA\ band 
centered at 1650~\AA\ is due to the B star.  We shall examine the effects 
of our assumptions shortly.  Only the \iue\/ fluxes between 1250 and 
1700\AA\ are incorporated into the fit of the SED, which constrains the 
physical properties of the B star.  This extends slightly beyond the 
1650\AA\ limit used for the STIS data, but recall that the \iue\/ data 
were obtained when the Cepheid was near minimum light, and nearly a factor 
of two fainter in the UV (see, Fig.~\ref{fluxes}).  

The parameters determined from the fit are given in Table~\ref{parameters}, 
where parameters that were fixed in the fit are enclosed in parentheses.  
Figure~\ref{rlam_no_mods} shows our fits to the centroids.  The points are 
the observed data and the solid curves are the fits obtained simultaneously 
with the fit to the fluxes.  The effects of spectral features on the 
centroids are clearly seen.  Figure~\ref{flam_no_mods} shows the fit to 
the SED below 1650\AA.  We do not show the fit to the binary SED longward 
of 1650\AA\ since it is, by definition, exact.  The extinction curve 
derived from the best fit is also shown in Figure~\ref{flam_no_mods}, where 
it is compared to a standard $R_V = 3.1$ curve from Fitzpatrick (1999).  

\begin{table}
\begin{center}
\caption{Parameter Values\label{parameters}}
\begin{tabular}{lr|lr}
\hline
Parameter       & Value      & Parameter       & Value      \\ \hline
$\Delta y_1$    &  $-0.010$  & $c_3$           &   4.13     \\
$\Delta y_2$    &   0.279    & $c_4$           &   0.82     \\
$\Delta y_3$    &  $-0.269$  & $\gamma$        &   0.9686   \\
T$_{\rm eff}^p$ &  [6297]    & $V_s$           &  (11.084)  \\ 
T$_{\rm eff}^s$ &  15735     & $(B-V)_0^s$     & ($-0.156$) \\
$\log g_p$      &  (4.00)    & $(U-B)_0^s$     & ($-0.597$) \\
$\log g_s$      &  [1.60]    & $V_p$           &  (7.362)   \\
$[{\rm m/H}]_p$ &  [0.00]    & $(B-V)_0^p$     &  (0.494)   \\
$[{\rm m/H}]_s$ &  -0.20     & $(U-B)_0^p$     &  (0.359)   \\
$E(B-V)$        &   0.53     & $\Delta \log L$ &  (0.95)    \\
$R(V)$          &   3.11     &                 &            \\ \hline
\end{tabular}
\end{center}

\medskip
Values in parenthesis were not involved in the fitting 
procedure. Values in square brackets were determined from a 
fit to the Cepheid SED derived from the initial fit.
\end{table}

%%%%%%%%%%%%%%%%%%%%%%%%%%%%%%% Figure 7 %%%%%%%%%%%%%%%%%%%%%%%%%%%%%%%%%
\begin{figure}
\centerline{\hbox{
\epsfxsize=3.5truein
\epsffile{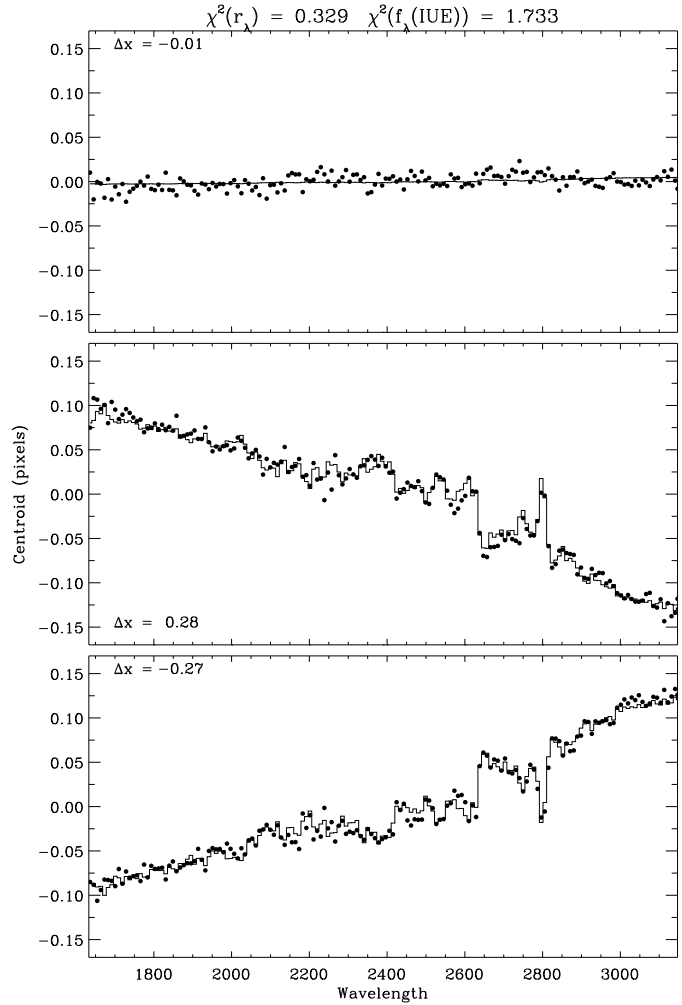}
}}
\caption{Fits to the mean centroids at each roll angle for AW Per.  Each 
mean centroid was fit simultaneously with the corresponding fluxes, optical 
photometry and interstellar extinction. A Kurucz model was used to fit the 
B star component, and the Cepheid flux was taken to be the difference 
between the reddened B star model and the observed flux.}
\label{rlam_no_mods}
\end{figure}
%%%%%%%%%%%%%%%%%%%%%%%%%%%%%%%%%%%%%%%%%%%%%%%%%%%%%%%%%%%%%%%%%%%%%%%%%

%%%%%%%%%%%%%%%%%%%%%%%%%%%%%%% Figure 8 %%%%%%%%%%%%%%%%%%%%%%%%%%%%%%%%%
\begin{figure}
\centerline{\hbox{
\epsfxsize=3.5truein
\epsffile{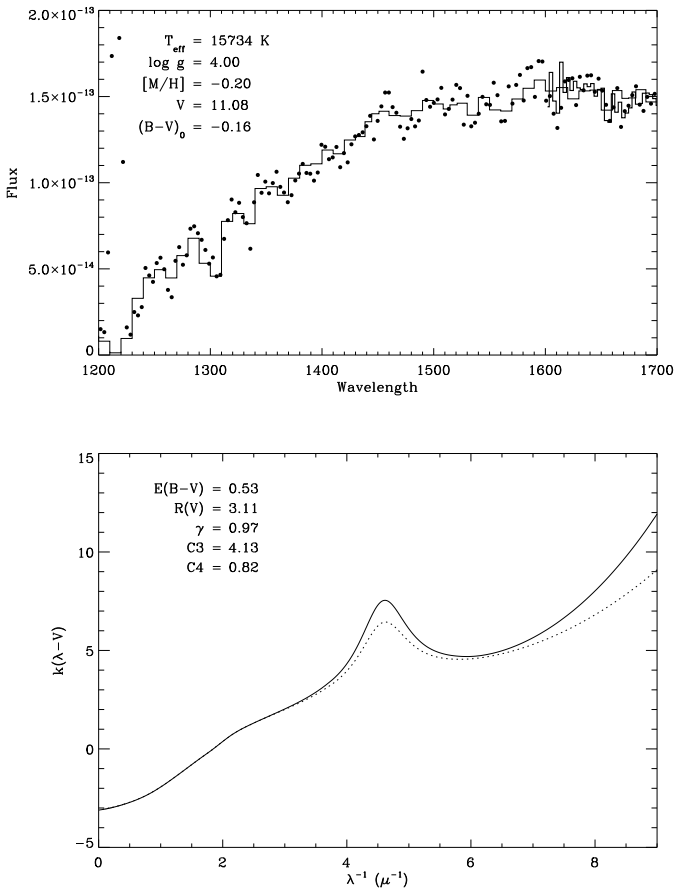}
}}
\caption{Top: Best fit B star (thin curve) compared to the \iue\/ (points) 
and STIS (thick curve) fluxes.  The model includes reddening.  We only show 
the far-UV region, since the fit is, by definition, exact for wavelengths 
longward of 1650\AA.  Bottom: AW~Per extinction curve determined by the 
simultaneous fit of the flux and centroids (solid curve) compared to a 
standard $R_V = 3.1$ curve (dotted) from Fitzpatrick (1999).}
\label{flam_no_mods}
\end{figure}
%%%%%%%%%%%%%%%%%%%%%%%%%%%%%%%%%%%%%%%%%%%%%%%%%%%%%%%%%%%%%%%%%%%%%%%%%

We can also estimate the physical parameters of the Cepheid component of 
the binary by fitting its mean SED inferred from fit.  This SED is found 
by subtracting the reddened B star model from the observed SED of the 
system and then correcting this difference for the effects of extinction.  
The unreddened SED plus its $V$, $(B-V)_0$ and $(U-B)_0$ (also inferred 
from the fit) were then fit to an Atlas~9 model.  The $V$, $(B-V)$ and 
$(U-B)$ photometry were initially assigned errors of 0.02, 0.01 and 
0.02~mag, respectively.  In performing this fit, we fixed the 
micro-turbulent velocity at 2~\kms, and allowed $T_{\rm eff}^p$ (the 
effective temperature of the primary), $\log g_p$ (the surface gravity of 
the primary) and ${[\rm m/H]}_p$ (the abundance of the primary), to vary.  
We had to restrict the surface gravity to be larger than 1.6, otherwise the 
fitting routine would seek $\log g_p$ values that were unrealistically 
small (we expect a $\log g_p \simeq 2.0$, e.g., Evans 1994).  Furthermore, 
we had to increase the weight (decrease the error) of the $(B-V)$ 
photometry by a factor of 10 in order to obtain reasonable agreement with 
the photometry.  Figure~\ref{cep_no_mods} compares the unreddened SED of 
the Cepheid to the best fit model.  The parameters derived from the fit are 
also listed in Table~\ref{parameters} and are enclosed in square brackets, 
to distinguish them from the parameters derived from the initial fit to the 
data.

%%%%%%%%%%%%%%%%%%%%%%%%%%%%%%% Figure 9 %%%%%%%%%%%%%%%%%%%%%%%%%%%%%%%%%
\begin{figure}
\centerline{\hbox{
\epsfxsize=3.5truein
\epsffile{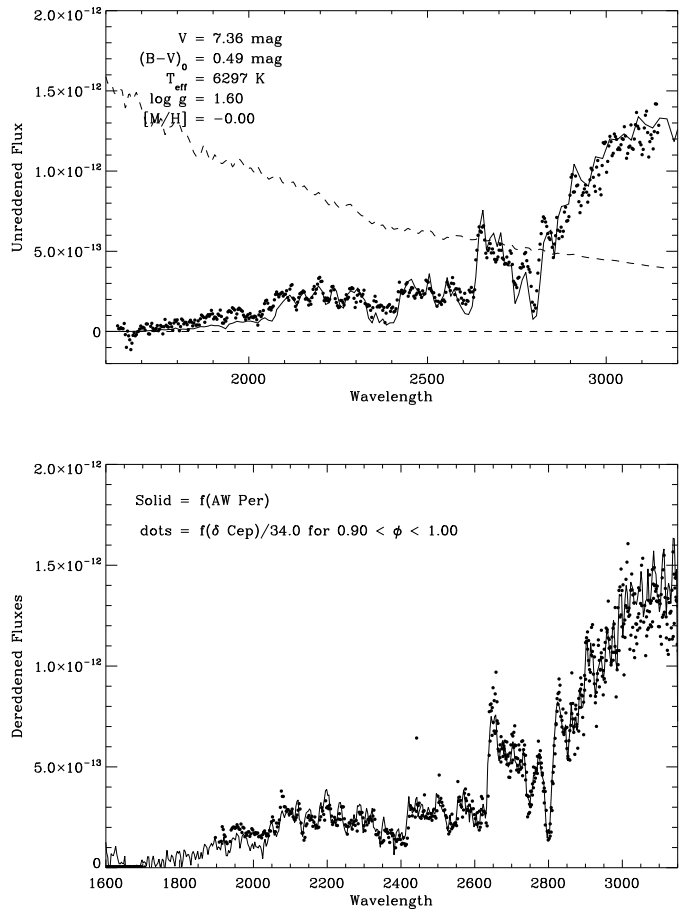}
}}
\caption{Top: Inferred dereddened Cepheid SED (points) compared to the 
best fitting Kurucz model (solid) and the dereddened flux of the best fit B 
star (dashed).  Bottom: Comparison of the unreddened Cepheid flux (solid 
curve) and an unreddened \iue\/ spectrum (dots) of $\delta$~Cep 
observations for $0.90 \leq \Phi \leq 0.95$.  The $\delta$~Cep flux is 
scaled by the difference between $V = 3.54$ at $\Phi = 0.925$ for 
$\delta$~Cep and $V = 7.37$, the magnitude of the primary in AW~Per at 
$\Phi = 0.92$ (the mean phase of the STIS observations).  As discussed in 
the text, the $\delta$~Cep spectrum is a combination of several \iue\/ 
spectra.}
\label{cep_no_mods}
\end{figure}
%%%%%%%%%%%%%%%%%%%%%%%%%%%%%%%%%%%%%%%%%%%%%%%%%%%%%%%%%%%%%%%%%%%%%%%%%5

It is also possible to test the reasonableness of the inferred UV Cepheid 
SED by comparing it to \iue\/ observations of the single Cepheid star 
$\delta$~Cep.  $\delta$~Cep has a period of 5.4~days, compared to 6.5~days 
for AW~Per, and its mean unreddened color is $\langle(B-V)\rangle = 0.57$.  
To obtain the intrinsic color of AW~Per, we use our derived color excess 
for the system and the intrinsic colors of the B star secondary from 
Table~\ref{parameters} and the mean magnitude of the system, $\langle V 
\rangle = 7.49$~mag, to correct the observed mean color of the system,
$\langle (B-V) \rangle = 1.06$~mag, for both extinction and the presence 
of the companion.  The result is $\langle (B-V)_0^p\rangle = 0.57$, 
identical to that of $\delta$~Cep (recall that the intrinsic color we 
derive for AW~Per is at $\Phi \simeq 0.92$).  Thus, the comparison between 
these two stars is expected to be quite good.  The bottom plot in 
Figure~\ref{cep_no_mods} compares the unreddened \iue\/ data (points) for 
$\delta$~Cep from several exposures obtained for $0.9 \leq \Phi(\delta 
{\rm Cep}) \leq 1.0$ to the unreddened Cepheid STIS spectrum (solid curve) 
of AW~Per.  Several \iue\/ exposures are required to produce the 
$\delta$~Cep spectrum since the dynamic range of \iue\/ was so limited and 
the range of the UV SED of $\delta$~Cep is so large.  The \iue\/ data had 
the Massa \& Fitzpatrick (2000) corrections applied, were dereddened by an 
$E(B-V) = 0.09$ (Dean et al.\ 1987) and scaled by $10^{-0.4 (7.37-3.54)}$, 
which corresponds to magnitude difference of AW~Per at $\Phi = 0.92$ (the 
mean for the STIS data) and $\delta$~Cep at $\Phi = 0.95$ (the mean of 
the \iue\/ data).  

Finally, we utilize the $\Delta y^{(n)}$ which resulted from the fits to 
derive the separation of the system and its position angle on the sky.  
These are found by fitting eq.~(\ref{solve}) to the plot of $\Delta y$ 
versus roll angle shown in Figure~\ref{result}.  The error bars at each 
orientation are the quadratic mean errors for that roll determined from 
the dispersion in the fits to the three individual sets of observations 
obtained at each orientation (see, next section).  The inverse of the 
errors squared were used to weight the fit. The final result of the 
analysis is a separation of $\theta = 13.74 \pm 0.26$~mas and a position 
angle $\phi = 184.16 \pm 1.94$~deg, for an accuracy of $\sim 2$\%.  

%%%%%%%%%%%%%%%%%%%%%%%%%%%%%%% Figure 10 %%%%%%%%%%%%%%%%%%%%%%%%%%%%%%%%%
\begin{figure}
\centerline{\hbox{
\epsfxsize=3.5truein
\epsffile{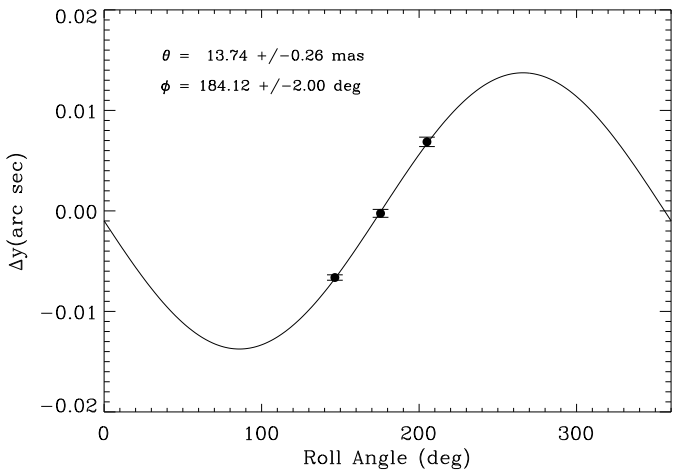}
}}
\caption{Determination of the angular separation of AW Per.  The 
observational errors for $\Delta y$ were determined from individual fits 
to the 3 independent offset observations at each roll angle.}
\label{result} 
\end{figure}
%%%%%%%%%%%%%%%%%%%%%%%%%%%%%%%%%%%%%%%%%%%%%%%%%%%%%%%%%%%%%%%%%%%%%%%%%

\subsection{Errors in the parameters}

In this section, we describe the internal, random, errors affecting our 
parameter determinations, and also examine the influence of systematic 
effects upon the results.  

The random errors were evaluated in two, independent ways.  One is the 
error estimates calculated by the least squares routine, which are 
determined by evaluating derivatives of the model.  These errors are 
listed in the second column of Table~\ref{errors}.  We also obtained error 
estimates by fitting the sets of observations obtained at the same off-set 
at each roll angle, independently.  These provide 3 sets of independent
observations and we used the parameters determined from each set to obtain 
standard deviations (S.D.s) of the model parameters.  These estimates 
(divided by $\sqrt 3$ applicable to the error in the mean) are listed in 
the third column of Table~\ref{errors}.  Notice that the errors in the 
$\Delta y^{(n)}$ determined from the S.D.s are nearly twice as large.  
To be conservative, these errors were used as the error shown in 
Figure~\ref{result} and in determining the errors in $\theta$ and $\phi$. 

\begin{table*}
 \begin{minipage}{120mm}
  \caption{Errors\label{errors}}
\begin{tabular}{lrr|rrr}
\hline
Param. & Prog. & S.D. & 
$|\delta \log g|$  & 
$|\delta \frac{f_{IUE}}{f_{{\rm STIS}}}|$ & 
$|\delta \frac{f_s}{f_P +f_s}|$ \\            \hline
$\Delta y^{(1)}$  & 0.004 & 0.015  & $1.4\times 10^{-4}$ & $ 6.5\times 10^{-6}$ & $5.0\times 10^{-4}$\\
$\Delta y^{(2)}$  & 0.005 & 0.017  &            $0.0019$ & $ 1.4\times 10^{-4}$ & $0.015$\\
$\Delta y^{(3)}$  & 0.005 & 0.014  &            $0.0021$ & $ 1.3\times 10^{-4}$ & $0.015$\\
T$_{\rm eff}^s$   &   248 &   105  &             $1205$  & 9.1                  & $37$\\
${\rm [m/H]}_s$   & 0.057 & 0.025  & $7.5 \times 10^{-5}$& $ 0.0016$            & $0.0064$\\
$E(B-V)$          & 0.001 & 0.038  &            $0.018$  & $ 0.0026$            & $0038$\\
$R_V$             & 0.031 &  0.12  &               0.11  & $ 0.0090$            & $2.7\times 10^{-4}$\\
$\gamma$          & 0.015 & $2.4 \times 10^{-4}$& $0.019$& $ 8.6\times 10^{-4}$ & $0.0025$\\
$c_3$             &  0.14 & 0.32   &           $0.049$   &                0.029 & 0.0055\\
$c_4$             & 0.019 & 0.066  &             0.014   & $ 4.3\times 10^{-3}$ & 0.0068\\
\hline
\end{tabular}
\end{minipage}
\end{table*}

Beside the random (or measurement) errors, systematic effects will also 
be present.  We characterize these by varying the different assumptions 
which enter the fitting procedure, and then examining their influence on 
the result.  To begin, we varied the assumed value of $\log g$ used to fit 
the B star by $\pm 0.5$, which should encompass all plausible values.  The 
result (the difference divided by 2) is listed in column~4 of 
Table~\ref{errors}.  Next, we tested the affect of assuming that the STIS 
(and not the \iue) fluxes are correct and allowed for the possibility that 
the B star accounts for only 95\%, instead of 100\% of the flux at 
1650\AA.  These results are listed in the last two columns of 
Table~\ref{errors}

As can be seen from Table~\ref{errors}, the varying the $\log g$ can cause 
a significant change in ${\rm T_{eff}}^s$, but has little effect on the 
$\Delta y^{(n)}$, which are the object of our analysis.  In fact, the only 
significant change in the $\Delta y^{(n)}$ result from our inability to 
determine whether the STIS or \iue\/ fluxes are correct, and even these 
errors are only of the same order of the errors determined from the 
repeated observations.  As a result, we conclude that the angular 
separation determined from our analysis is very robust to variations in 
the assumptions or input parameters.  

\section{DISCUSSION}\label{discussion} 

We have seen that the separation determined from the fit is quite stable.  
We now discuss the physical parameters determined from our fits 
(Table~\ref{parameters}), their reliability and their implications.  

We first consider the Cepheid SED derived from the fit.  It is compared to 
the best fitting Atlas~9 model in top panel of Figure~\ref{cep_no_mods}.  
This ``best fitting'' model is not a very good fit, since it lies 
systematically below the observed flux in far-UV flux and over it in the 
near-UV flux.  Furthermore, the agreement with the optical photometry is 
not very good.  The model predicts $V = 7.362$, $(B-V) = 0.470$ and $(U-B) 
= 0.309$.  The the agreement with the $(B-V)$ color given in 
Table~\ref{parameters} is fair, but recall that it was given a very 
large weight.  The agreement with the inferred $(U-B)$ is not very good at 
all.  The poor overall fit probably results from the short comings of 
Atlas~9 models for Cepheids discussed below.  

The bottom panel of Figure~\ref{cep_no_mods} compares the unreddened SED 
of the Cepheid component of AW~Per to the unreddened SED of the single 
Cepheid, $\delta$~Cep at approximately the same phase.  
This figure demonstrates three points.  First, the two SEDs agree 
surprising well.  Second, the strong far-UV flux in the derived SED 
relative to the models is also present (and slightly larger) in 
$\delta$~Cep, so the derived SED is quite reasonable.  Third, the flux in 
$\delta$~Cep is extremely small for wavelengths shortward of 1650\AA, 
bolstering our assumption that all of the flux in AW~Per observed below 
1650\AA\ is due to the B star secondary.  

So, why is the Atlas~9 model fit of the Cepheid so poor?  One must remember 
that Cepheid UV SEDs depend on numerous, ill-defined physical processes 
that are not fully incorporated into the Atlas~9 models.  These include 
spherical extension, which can enhance the UV flux from an atmosphere 
(see Fig.~4 in Hauschildt et al.\ 1999), chromospheres (e.g., Sasselov \& 
Lester 1994), the amount of convective energy transport (Castelli, 
Gratton, \& Kurucz 1997) and the details of the line blanketing (Prieto, 
Hubeny, \& Lambert, 2003).  In addition, there are inevitably dynamical 
effects that are not treated by the models.  

In fact, we initially attempted to fit the data with using an approach that 
employed models for both the Cepheid and the B star.  However, we abandoned 
it because it produced poor fits and the separations that were $\sim 10$\% 
larger than those derived from the adopted technique.  The origin of the 
systematic difference in the centroids can be traced to the gradient 
in the flux residuals seen in the top of Figure~\ref{cep_no_mods}.  These 
propagate into the fits of the centroids.  Perhaps the use of more detailed 
Cepheid models could solve this problem.  
 
In spite of these difficulties, it is of interest to examine the physical 
parameters determined from the Cepheid model.  To begin, \teff\/ of the 
best fit model agrees reasonably well with previous estimates for Cepheid 
temperatures near maximum light (Evans \& Teays 1996, Fry \& Carney 1999, 
Kovtyukh \& Gorlova 2000).  On the other hand, the fit selects a very low 
surface gravity and would have settled on an even lower value if it had 
been allowed to do so.  It is also interesting that the Cepheid model has 
a significantly different metallicity than the B star.  However, this may 
not be too strange.  Instead, it may simply reflect the fact that the 
[m/H] parameter in cooler models responds more to spectral features 
produced by CNO elements, while the same parameter in the B stars responds 
to the Fe abundance (Fitzpatrick \& Massa 1999).  

Next, we consider the parameters determined for the B star.  The model fit 
to the far-UV (Fig.~\ref{flam_no_mods}) is quite good, and the extinction 
curve, while distinctly different from the canonical $R_V = 3.1$ curve, is 
rather unremarkable, with parameters well within normal bounds (e.g., 
Fitzpatrick \& Massa 1990, Valencic et al.\ 2004).  Also, the [m/H] for 
the B star is well within the expected range for such stars (e.g., 
Fitzpatrick \& Massa 1999, 2005) and the inferred color excess is quite 
close to previous determinations (Evans 1994).  It should not be surprising 
that these fits are so good, since both the extinction model and the 
ability of the Atlas~9 models to describe normal B star spectra are well 
documented.  Notice that \teff\/ we derive is considerably hotter than 
previously estimated by Evans (1994), and lies somewhat closer to the ZAMS 
(see, Fig.~7 in Evans 1994).  However, its probable mass, $\sim 5M_\odot$ 
(based on its \teff, Andersen, 1991), remains significantly less than the 
lower limit of $\sim 6.6M_\odot$ determined from the radial velocity orbit 
of the primary by Evans et al.\ (2000).  Thus, it still appears likely that 
the B star component of AW~Per must also be a binary.  

\section{SUMMARY}\label{summary} 

We have shown that the signatures of the Cepheid and B star components 
of AW Per are clearly present in the wavelength dependence of the 
centroid of its spectrum.  This result demonstrates the power of our 
approach.  A simple model was devised to extract the angular separation 
of the binary from the centroid measurements.  The accuracy of the 
angular separation is $\sim 2$\%, or $\pm$~a few $\times 10^{-4}$\arcsec!   
We also demonstrated that the results are extremely stable to variations 
in the expected systematic effects in the data and its analysis.  We also 
showed that one possible source of uncertainty in the current data is the 
absolute level of the far-UV data.  Higher quality far-UV observations to 
secure the B star flux level and secure its parameters would be extremely 
useful.  

Our final results are listed in Table~\ref{parameters}.  In addition to 
the angular separations and position angle, these include a Cepheid 
temperature and systemic extinction that agree with previous estimates 
and a B star secondary temperature that is considerably hotter than 
previously thought (e.g., Evans, 1994).  However, the likely mass of the 
secondary still appears too small to account for the minimum mass of the 
secondary inferred by the radial velocity of the primary.  Consequently, 
it is likely that the B star component of AW~Per is also be a binary.  

Finally, the long period of AW~Per's orbit means that it will be a few 
years before the separation changes enough for the second independent 
observation needed to determine $\sin i$ can be obtained.  

\section*{Acknowledgments}

We would like to thank Karla Peterson and Charles Proffit of STScI, who 
provided valuable guidance in preparing the observations.  This work was 
supported by NASA grants to SGT, Inc. and SAO. % through grant NAG-xxxxx to SGT Inc and NAG-xxxxx to SAO.  

\appendix

\section{MATHEMATICAL DETAILS}
This appendix provides a detailed derivation of how the wavelength 
dependence of the centroid of the a dispersed image can be used to 
determine the separation of a binary whose components have different 
colors.

Consider the set of angular coordinates $x$ and $y$ which are parallel 
and perpendicular to the dispersion, respectively, with $x$ increasing in 
the direction of increasing wavelength (this is the standard STIS 
coordinate system, Kim et al.\ 2003).  Now, let $h(y)$ be the 
instrumental profile in the cross-dispersion direction, $y$. Then the 
spectrum of a single star located at $y = y_0$ can be expressed as
\begin{equation}
\label{one}f(\lambda ,y)=N(\lambda )h(y-y_0)
\end{equation}
where $\lambda = \lambda(x)$, and $N(\lambda)$ is the photon flux at 
$\lambda$ (we assume infinite resolution in the wavelength direction).

If the spectra of the primary and secondary components of the binary are 
centered at $y_p$, and $y_s$, then their spectra are separated by $\Delta 
y = y_s -y_p$, and the image of the binary spectrum is given by 
\begin{equation}
\label{two}f(\lambda,y)=N_p(\lambda)h(y-y_p)+N_s(\lambda)h(y-y_p-\Delta y).
\end{equation}
If $\Delta y$ is small compared to structure in $h(y)$, this equation can 
be approximated by
\begin{eqnarray}
f(\lambda,y) & \simeq & N_p(\lambda)h(y-y_p) \nonumber\\
             &    & +N_s(\lambda)\left[h(y-y_p)+\Delta y 
                   \frac{dh(y)}{dy}\bigg\vert_{y=y-y_p}\right]\nonumber  \\
             & =  & [N_p(\lambda)+N_s(\lambda)] \times \nonumber \\ 
             &    & \left[h(y-y_p) + \frac{N_s(\lambda)}{N_p(\lambda)
                    +N_s(\lambda)}\Delta y \frac{dh(y)}{dy} \nonumber  
                    \bigg\vert_{y=y-y_p}\right] \\
         & \simeq & [N_p(\lambda)+N_s(\lambda)] \times \nonumber \\ 
               &  & h\left[y-\left(y_p+ \frac{\Delta y N_s(\lambda)}
               {N_p(\lambda)+N_s(\lambda)}\right) \right]
\end{eqnarray}
Therefore, the wavelength dependence of the centroid of the spectrum will
vary as
\begin{equation}
\label{final}y(\lambda)=y_p+\frac{\Delta y N_s(\lambda)}{N_p(\lambda) +
N_s(\lambda)} = y_p+\frac{\Delta y}{1+R(\lambda)}
\end{equation}
where $R(\lambda)=N_p(\lambda)/N_s(\lambda)$ is the flux ratio of the binary
components.

Now, the separation $\Delta y$ depends on both the separation of the binary, 
$\theta$, and the orientation of the system relative to the dispersion 
direction.  The position angle on the sky of the binary, $\phi$, is 
defined as the angle measured the c.c.\ from north to east, with the primary 
at the origin.  The angle $\alpha^{(n)}$ (also measured the c.c.\ from north 
to east) is defined as the angle of a line in the dispersion direction
pointing in the direction of increasing wavelength for the $n^{th}$ 
telescope orientation.  In this case, $\phi -\alpha^{(n)}$ is the angle 
between the dispersion and a line connecting the binary components and 
$\Delta y^{(n)}  = \theta \sin(\phi - \alpha^{(n)})$ is the displacement of 
the two spectra of the binary (note that when $\phi - \alpha = 0$, $\pm 
180^\circ$, $\Delta y = 0$).  Therefore, the observation obtained with the
telescope in the $n^{th}$ orientation can be expressed as
\begin{equation}
y(\lambda)^{(n)} = y_p^{(n)}+\theta \sin(\phi -\alpha^{(n)}) [1 +
R(\lambda)]^{-1}
\end{equation}
where $y_p^{(n)}$ is the wavelength {\em independent} displacement of the
$n^{th}$ exposure in $y$.  

To extract both $\theta$ and $\phi$ from the observed centroids, at 
least two observations at different $\alpha$'s are required.  Therefore, as 
long as the {\em relative} fluxes of the binary components are known, a 
linear regression of the wavelength dependence of the centroid against 
$[1 +R(\lambda)]^{-1}$ gives $\Delta y^{(n)}$ for that observation. The 
$y_p^{(n)}$ are constant terms related to the absolute position of the 
primary star, although in practice the they cannot be reliably disentangled 
from the large random errors in the absolute position of the binary on the 
detector at each orientation.  

Once the $\Delta y^{(n)}$ are determined for each orientation, these are 
plotted against the known quantities, $\alpha^{(n)}$.  Since 
\begin{equation}
\Delta y^{(n)} = \theta \sin(\phi -\alpha^{(n)}) 
\end{equation}
fitting a sine function to the $\Delta y^{(n)}$ as a function of the 
$\alpha^{(n)}$ determines $\phi$ and $\theta$, the observables of an 
astrometric binary at the epoch of the observations.


\begin{thebibliography}{99}

\bibitem[\protect\citeauthoryear{Andersen}{1991}]{a91} Andersen, J. 1991, 
  A\&A Rev, 3, 91

\bibitem[\protect\citeauthoryear{Beckers}{1982}]{b82} Beckers, J.M. 1982, 
  Opt. Acta, 29, 361

\bibitem[\protect\citeauthoryear{Benedict et al.}{2002}]{b2002} Benedict, 
  G.F., McArthur, B.E., Fredrick, L. W., et al.\ 2002 AJ, 124, 1695

\bibitem[Casstelli et al. (1997)]{c97} Castelli, F., Gratton, R.G. \& 
  Kurucz, R.L. 1997, A\&A,  318, 841

\bibitem[\protect\citeauthoryear{Dean et al.}{1987}]{d87} Dean, J.F., 
  Warren, P.R. \& Cousins, A.W.J. 1987, MNRAS, 183, 569

\bibitem[\protect\citeauthoryear{Evans}{1989}]{e89} Evans, N.R. 1989, AJ, 
  97, 1737

\bibitem[\protect\citeauthoryear{Evans}{1994}]{e94} Evans, N.R. 1994, ApJ, 
  436, 273

\bibitem[\protect\citeauthoryear{Evans}{1995}]{e95} Evans, N.R. 1995, ApJ, 
  445, 393 

\bibitem[\protect\citeauthoryear{Evans \& Teays}{1996}]{et96} Evans, N.R. 
  \& Teays, T.J. 1996, AJ, 112, 761

\bibitem[\protect\citeauthoryear{Evans et al.}{1998}]{e98} Evans, N.R., 
  Bohm-Vitense, E., Carpenter, K., Beck-Winchatz, B. \& Robinson, R. 1998, 
  ApJ, 494, 768

\bibitem[\protect\citeauthoryear{Evans et al.}{2000}]{e2000} Evans, N.R., 
  Vinko, J. \& Wahlgren, G.M. 2000, AJ, 120, 407

\bibitem[\protect\citeauthoryear{Evans et al.}{2007}]{e2007} Evans, N. R., 
  et al.\ 2007, in Hartkopf, W.I. Guinan, E.F. \& Harmanec, P. eds., 
  ``Binary Stars as Critical Tools and Tests in Contemporary 
  Astrophysics'', IAU. Symp 240, 102

\bibitem[\protect\citeauthoryear{Fitzpatrick}{1999}]{f99} Fitzpatrick. E.L. 
  1999, PASP, 111, 63

\bibitem[\protect\citeauthoryear{Fitzpatrick \& Massa}{1990}]{fm90} 
  Fitzpatrick, E.L. \& Massa, D. 1990, ApJS, 72, 163

\bibitem[\protect\citeauthoryear{Fitzpatrick \& Massa}{1999}]{fm99} 
  Fitzpatrick, E.L. \& Massa, D. 1999, ApJ, 525, 1011

\bibitem[\protect\citeauthoryear{Fitzpatrick \& Massa}{2005}]{fm05} 
  Fitzpatrick, E.L. \& Massa, D. 2005, AJ, 129, 1642

\bibitem[\protect\citeauthoryear{Fry \& Carney}{1999}]{fk99} Fry, A.M. \& 
  Carney, B.W. 1999, AJ, 118, 1806

\bibitem[\protect\citeauthoryear{Groenewegen \& Oudmaijer}{2005}]{go05} 
  Groenewegen, M.A.T. \& Oudmaijer R.D. 2000, A\&A, 356, 849	

\bibitem[\protect\citeauthoryear{}{}]{} Hauschildt, P.H., Allard, F., 
  Ferguson, J., Baron, E. \& Alexander, D.R. 1999, ApJ, 525, 871

\bibitem[\protect\citeauthoryear{}{}]{} Kervella, P., Fouqu{\'e}, P., 
  Storm, J., Gieren, W.P., Bersier, D., Mourard, D., Nardetto, N., Foresto, 
  V. 2004, ApJ, 604, L113

\bibitem[\protect\citeauthoryear{Kovtyukh \& Gorlova}{2000}]{kg00} 
  Kovtyukh, V.V. \& Gorlova, N.I. 2000, A\&A, 358, 587

\bibitem[\protect\citeauthoryear{Kriss}{1998}]{k98} Kiss, L.L. 1998, MNRAS, 
  297, 825

\bibitem[\protect\citeauthoryear{Kurucz}{1991}]{atlas} Kurucz, R.L. 1991, 
  in ``Stellar Atmospheres: Beyond Classical Models'', ed. L. Crivellari, 
  I. Hubeny, \& D.G. Hummer (NATO ASI Ser. C.; Dordrect: Kluwer), 441

\bibitem[\protect\citeauthoryear{Lindler}{1998}]{lindler} Lindler, D. 1998, 
  CALSTIS Reference Guide (CALSTIS Version 5.1) (Greenbelt, MD:GSFC)

\bibitem[\protect\citeauthoryear{Massa \& Endal}{1987}]{me1} Massa, D. \& 
  Endal, A.S. 1987, AJ, 93, 760

\bibitem[\protect\citeauthoryear{Massa \& Fitzpatrick}{2000}]{mf00} Massa, 
  D. \& Fitzpatrick, E. L. 2000, ApJS, 126, 517

\bibitem[\protect\citeauthoryear{Moffett \& Barnes}{1984}]{mb94} Moffett, 
  T.J., \& Barnes, T.G. 1984, ApJS, 55, 389

\bibitem[\protect\citeauthoryear{Nichols \& Linsky}{1996}]{nl96} Nichols, 
  J.S. \& Linsky, J.L. 1996, AJ, 111, 517

\bibitem[\protect\citeauthoryear{Preito et al.}{2003}]{p03} Prieto, C.A. , 
  Hubeny, I. \& Lambert, D.L. 2003, ApJ, 591, 1192

\bibitem[\protect\citeauthoryear{Porter et al.}{2004}]{p04} Porter, J.M., 
  Oudmaijer, R.D.\& Baines, D. 2004, A\&A, 428, 327

\bibitem[\protect\citeauthoryear{STIS}{2003}]{q03} Kim Quijano, J., et al. 
  2003, ``STIS Instrument Handbook'', Version 7.0, (Baltimore: STScI).

\bibitem[\protect\citeauthoryear{Romaniello et al.}{2005}]{r05} Romaniello, 
  M., Primas, F., Mottini, M., Groenewegen, M., Bono, G. \& François, P. 
  2005, A\&A, 429L, 37

\bibitem[\protect\citeauthoryear{Sasselov, D.D. \& Lester}{1994}]{sl94} 
  Sasselov, D.D. \& Lester, J.B. 1994, ApJ, 423, 795

\bibitem[\protect\citeauthoryear{Schaller et al.} {1992}]{s92} Schaller, 
  G., Schaerer, D., Meynet, G., \& Maeder, A. 1992, A\&AS, 96, 269

\bibitem[\protect\citeauthoryear{Schmidt \& Parsons}{1982}]{sp82} Schmidt, 
  G.S. \& Parsons, S.B. 1982, ApJS, 48, 185

\bibitem[\protect\citeauthoryear{Simon}{1990}]{sim90} Simon, N.R. 1990 in 
  Cacciari, C. \& Clementini, G. eds., ASP Conf. Ser., 11, Confrontation 
  between stellar pulsation and evolution, p193

\bibitem[\protect\citeauthoryear{Szabados}{1980}]{sz80} Szabados, L. 1980, 
  Commun. Konkoly Obs. Hung. Acad. Sci., Budapest, No.76

\bibitem[\protect\citeauthoryear{Szabados}{1991}]{sz91} Szabados, L. 1991, 
  Commun. Konkoly. Obs. Hung. Acad. Sci., Budapest, No.96

\bibitem[\protect\citeauthoryear{Valencic et al.}{2004}]{v04} Valencic, 
  L.A., Clayton, G.C., \& Gordon, K.D.  2004, ApJ, 616, 912 

\end{thebibliography}
\end{document}